\documentclass[aps,prb,twocolumn,superscriptaddress, show pacs]{revtex4-1}
\usepackage{bm, amsmath}
\usepackage{graphicx}
\usepackage{color}
\usepackage{epstopdf}
\usepackage{graphicx}
\usepackage{dcolumn}
\usepackage{bm}
\usepackage{subfigure}

\begin{document}

\title{Theory of skyrmion states in liquid crystals: \\
axisymmetric cholesteric bubbles  (spherulites)}

\author{A. O. Leonov}
\thanks{Corresponding author, Email: a.leonov@rug.nl}
\affiliation{IFW Dresden, Postfach 270016, D-01171 Dresden, Germany}   
\affiliation{Zernike Institute for Advanced Materials, University of Groningen, Groningen, 9700AB, The Netherlands}

\author{I. E. Dragunov}
\affiliation{Donetsk Institute for Physics and Technology,
340114 Donetsk, Ukraine
}

\author{U.K. R\"o\ss ler}
\author{A.~N.~Bogdanov} \thanks{a.bogdanov@ifw-dresden.de }
\affiliation{IFW Dresden, Postfach 270016, D-01171 Dresden, Germany}

\date{\today}

\begin{abstract}
{Within the Oseen-Frank theory we derive numerically exact solutions 
for axisymmetric localized states in chiral liquid crystal layers with
homeotropic anchoring. These solutions describe  recently observed 
two-dimensional skyrmions in confinement-frustrated chiral nematics 
[P. J. Acherman et al. Phys. Rev. E \textbf{90}, 012505 (2014)].
We stress that these solitonic states arise due to a fundamental stabilization 
mechanism responsible for the formation of skyrmions in cubic helimagnets 
and other noncentrosymmetric condensed-matter systems.
}
\end{abstract}



\maketitle

\section{Introduction}

The concept of nonsingular multidimensional topological solitons
(commonly referred as \textit{skyrmions})
plays an important role in many branches of modern physics
\cite{skyrmion}. Particularly, specific skyrmionic states
can arise in condensed-matter systems with intrinsic 
and induced chirality \cite{JETPL1995,JETP1998}.
During last two decades regular solutions for such \textit{chiral}
skyrmions have been derived for different classes of noncentrosymmetric
magnets \cite{JMMM1994,Nature2006,Butenko2010,Borisov2010} 
and liquid crystals \cite{JETP1998,JETPL2000,Fukuda2011}.
In experiment, first indirect evidences for
the existence of multidimensional chiral modulations 
have been reported in the precursor region of a helimagnet MnSi 
(see e.g. contributions \cite{Muhlbauer2009} and a discussion 
and bibliography in \cite{Wilhelm2011}).
The following direct observations of chiral skyrmions in 
nanolayers of a cubic helimagnet (Fe$_{0.5}$Co$_{0.5}$)Si
\cite{Yu2010} have triggered intensive investigations of
these solitonic states in different classes of noncentrosymmetric magnets
(see e.g. \cite{Seki2012, Huang2012,Romming2013,JPCS2011}
and bibliography in \cite{Wilson2014}). 
Chiral magnetic skyrmions are now considered as promising objects 
for applications in magnetic data storage technologies and in the
emerging spin electronics
\cite{Romming2013,Kiselev2011,Fert2013}.

Recently two-dimensional axisymmetric localized strings 
analogous to isolated magnetic skyrmions have been optically generated 
in thin layers of a chiral nematic confined between substrates with
perpendicular (\textit{homeotropic}) surface anchoring 
\cite{Ackerman2014}. 
During last four decades a rich variety of  2D and 3D 
localized structures with different types of singularities
have been observed in confined chiral 
liquid crystals as \textit{cholesteric bubbles} 
(\textit{spherulitic domains}) \cite{Kawachi1974,Oswald2005},
cholesteric fingers \cite{Oswald2005,Smalyukh2005},
torons \cite{Smalyukh2010,Chen2013}, and other specific solitonic
textures \cite{Pandey2014}. 

Importantly that in most of nonlinear field models 
\textit{static} multidimensional solitons are unstable 
and collapse spontaneously under the influence of internal 
and external perturbations (this mathematical truth
is known in physicis of solitons as \textit{Derrick-Hobart
theorem} \cite{Derrick1964}).
However, in condensed-matter systems lacking inversion symmetry 
(such as magnets, multiferroics, ferroelectrics, 
and chiral liquid crystals) there exist 
specific interactions imposed by the handedness 
of the underlying structure which stabilize 2D and 3D isolated 
states \cite{JETPL1995,JMMM1994,JETPL2000}.
It was shown by direct calculations that this fundamental 
stabilization mechanism is responsible for the
formation of skyrmions (nonsingular solitons)
in bulk and confined chiral liquid crystals
\cite{JETPL2000,Fukuda2011}.
%

In this paper we apply the nonsingular model \cite{JETPL2000,Akahane1976} 
to investigate basic properties of two-dimensional axisymmetric skyrmions 
in thin layers of chiral liquid crystals. 
By numerically solving the differential equations
minimizing the Frank functional we derive the equilibrium structures
of confined chiral modulations (isolated and embedded skyrmions,
helicoids) as functions of the material parameters and applied
electric (or/and magnetic) fields.

\section{Isolated skyrmions}
Within the continuum theory the equilibrium distributions of the director 
$\mathbf{n} (\mathbf{r})$ in confined liquid crystals are derived by solving 
the Euler equations for the Frank free energy density functional \cite{Kleman2003}
\begin{align}
& f (\mathbf{n})  =\frac{\mathit{K}_1}{2}(\rm{div}\,\mathbf{n})^2
+\frac{\mathit{K}_2}{2}(\bf{n}\cdot\rm{rot}\,\mathbf{n}-q_0)^2
\nonumber\\
& +\frac{\mathit{K}_3}{2}(\mathbf{n}\times\rm{rot}\,\mathbf{n})^2
-\frac{\varepsilon_a}{2}(\mathbf{n}\cdot\mathbf{E})^2 
-\frac{ \chi_a}{2}(\mathbf{n}\cdot\mathbf{H})^2.
\label{Frank}
\end{align}
Here, $K_i\, (i=1,2,3)$ and $q_0$ are elastic constants; $\mathbf{E}$ and $\mathbf{H}$ are the vectors
of applied electric and magnetic  fields, and $\varepsilon_a$ and $\chi_a$ are values of dielectric and diamagnetic anisotropies.
Because dielectric and diamagnetic anisotropy energy contributions 
have the same functional form (\ref{Frank}) they lead to the same
solutions. For the sake of simplicity we  consider only effects 
imposed by applied electric fields.
We also  restrict our analysis by the one constant approximation 
($K_1 = K_2 = K_3 = K$).
\begin{figure}
\centering
\includegraphics[width=\columnwidth]{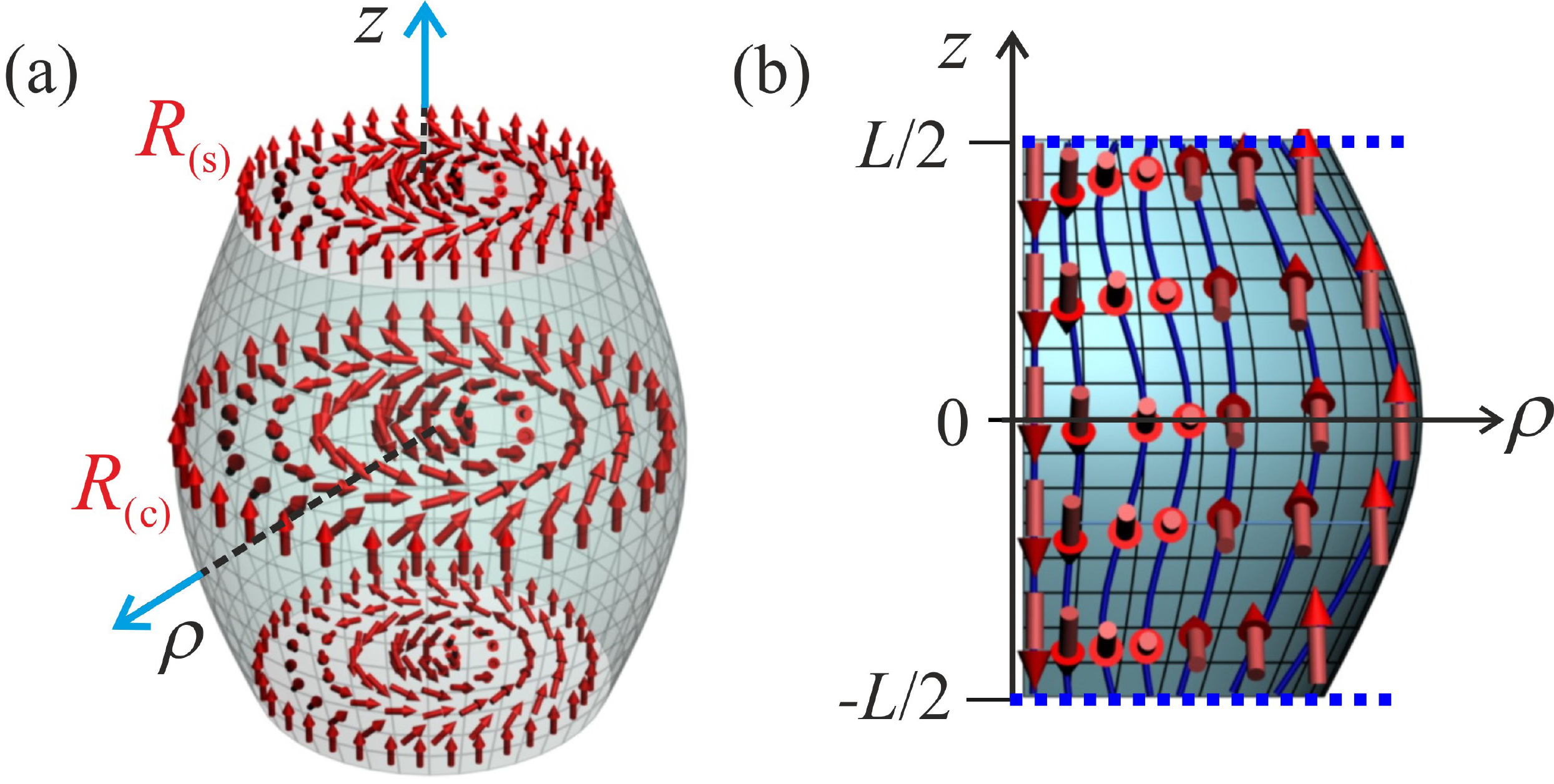}
\caption{ (Color online)
Axisymmetric skyrmion  in a chiral liquid crystal 
layer of thickness $L$ confined by
surfaces $z = \pm L/2$ with homeotropic anchoring.
The director $\mathbf{n}$ is designated by arrows 
to demonstrate a fixed rotation sense.  
(a) Due to homeotropic anchoring the cores
of the surface layers (\textit{s}) are smaller than in the
central (\textit{c}) layer.
(b) Cut in the $\rho z$ half-plane shows rotation of
$\mathbf{n}$ along the $\rho$ axes for different values
of $z$. Isoclines $\theta = \pi i/6$ ($ i = 0...6$) calculated
for solutions $\theta (\rho, z)$ in Fig. 2 are indicated
with solid (blue) lines.
Characteristic widths of the skyrmion core $R(z)$ are defined
by Eq. (\ref{width}).
\label{pics}
}
\end{figure}

The equilibrium states of an infinite chiral liquid crystal are characterized
by two material parameters
\begin{equation}
p = \frac{2 \pi}{q_0}, \quad  E_0 = \frac{\pi K q_0}{2} \sqrt{\frac{K }{\varepsilon_a}}
\label{pitch}
\end{equation}
where \textit{helix pitch} $p $ is the width of one complete turn 
($\Delta \theta = 2\pi$) of the director $\mathbf{n}$ along the 
helical axis and $E_0$ is a value of the applied field unwinding 
the helix into a set of isolated kinks \cite{Kleman2003}.

We consider a plate  infinite in $x$- and $y$-directions and confined by parallel 
planar surfaces at $z=\pm L/2$ as a model for a thin layer of a chiral nematic 
liquid crystal of thickness $L$ sandwiched between two glass plates. 
To investigate axisymmetric localized solutions we introduce cylindrical 
coordinates for the spatial variable $\mathbf{r}$ and spherical coordinates 
for the director $\mathbf{n}$:  
$\mathbf{r} = (\rho \cos \varphi, \rho \sin \varphi, z)$,
\begin{equation}
\mathbf{n} = (\sin \theta \cos \psi, \sin \theta \sin \psi, \cos \theta).
\label{coordinates}
\end{equation}
The total energy for  isolated solutions of type 
$ \theta = \theta (\rho, z)$, $\psi = \psi (\varphi)$
can be written as
\begin{align}
& F = \frac{K}{2}\int_0^{2\pi} d \varphi \int_{-L/2}^{L/2} d z \int_0^{\infty} \rho d\rho
\left[ w (\theta, \psi) + w_s (\theta) \right] 
\label{energy1}
\end{align}
where 
\begin{align}
w = &  \left( \frac{\partial \theta}{\partial z} \right)^2 +
\left( \frac{\partial \theta}{\partial \rho} \right)^2 
+ \frac{\sin^2 \theta}{\rho^2} \left( \frac{\partial \psi}{\partial \varphi} \right)^2
+\frac{ \varepsilon_a E^2}{K} \sin^2 \theta
\nonumber\\
& + 2 q_0 \left[ \left( \frac{\partial \theta}{\partial \rho} \right) + \frac{\sin \theta \cos \theta}{\rho}
 \left( \frac{\partial \psi}{\partial \varphi} \right) \right] \sin (\psi- \phi),
\label{density1}
\end{align}
the surface energy $w_s (\theta) =  (K_s/K)\sin^2 \theta  \delta (z \pm L/2) $ where
$\delta (x)$ is the Dirac function. For $K_s > 0$ the energy density $w_s (\theta)$
  decribes a \textit{homeotropic} anchoring.

\begin{figure}
\centering
\includegraphics[width=0.9\columnwidth]{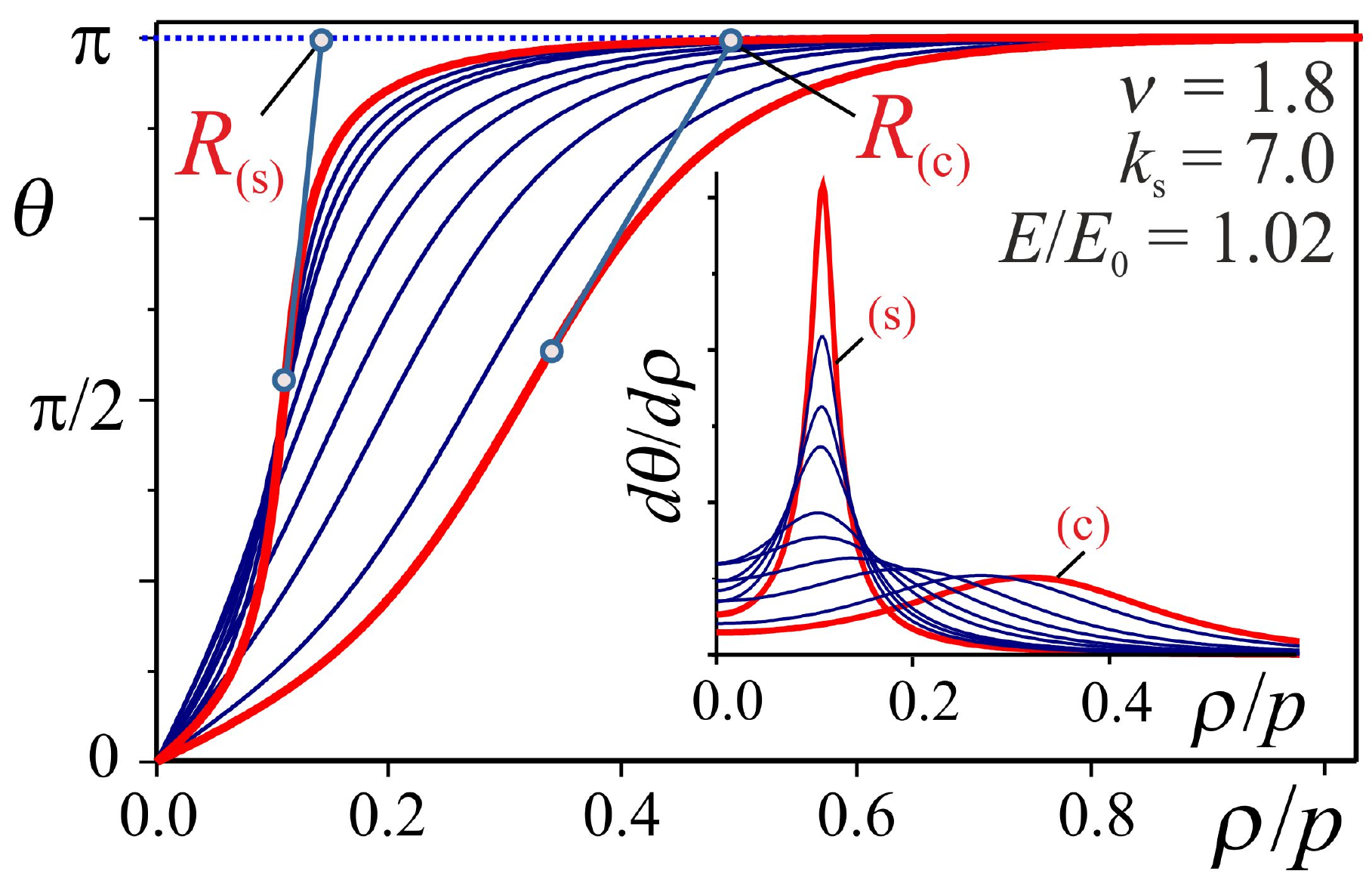}
\caption{ (Color online)
Solutions of Eqs. (\ref{EulerSp1}), (\ref{EulerSp2}) in a layer with confinement ratio
$\nu = L/p = 1.8$, surface anchoring $k_s = 7.0$, and the appplied
field $ E = 1.02 E_0$. Functions $\theta (\rho, z)$ are plotted as a set of profiles 
$\theta(\rho)$ for different values of $z$ ($0 < z < L/2$). Thick (red) lines
show solutions  in the center (c), $\theta (\rho, 0)$ and on the surfaces (s) 
of the layer, $\theta (\rho, \pm L/2)$.
The correspoding core sizes $R_{(c)}$,  $R_{(s)}$ are derived from Eq. (\ref{width}).
Inset shows corresponding profiles $d \theta / d \rho (\rho)$ 
indicating the inflection points of $\theta (\rho)$ profiles.
\label{profiles}
}
\end{figure}

Minimization of functional (\ref{energy1}) yields  $\psi = \varphi + \pi/2$, and
the equilibrium profiles $\theta (\rho, z)$ are derived by solving the Euler equation
\begin{align}
\frac{\partial^2 \theta}{\partial z^2}+\frac{1}{\rho} \frac{\partial \theta}{\partial \rho}
&+\frac{\partial^2 \theta}{\partial \rho^2} -\frac{1}{\rho^2}\sin \theta \cos \theta
\nonumber\\
&-\frac{2 q_0}{\rho} \sin^2 \theta-\frac{\varepsilon_a E^2}{K}\sin \theta \cos \theta=0,
\label{EulerSp1}
\end{align}
with boundary conditions $\theta(0,z)=\pi$, $\theta(\infty,z)=0$,
\begin{equation}
\left( \frac{\partial \theta}{\partial z}+\frac{K_s}{K}\sin\theta\cos\theta \right)|_{z=\pm L/2}=0.
\label{EulerSp2}
\end{equation}

The solutions $\theta (\rho, z)$ of Eqs. (\ref{EulerSp1}), (\ref{EulerSp2})
depend on the three control parameters
\begin{equation}
 E/E_0, \quad k_s = K_s/(K q_0), \quad \nu = L/p
\label{parameters}
\end{equation}
expressed as reduced values of the applied electric field ($E/E_0$), 
the homeotropic anchoring ($k_s$), and the layer thickness $\nu$ known
as \textit{confinement ratio}.

\begin{figure*}
\centering
\includegraphics[width=2.0\columnwidth]{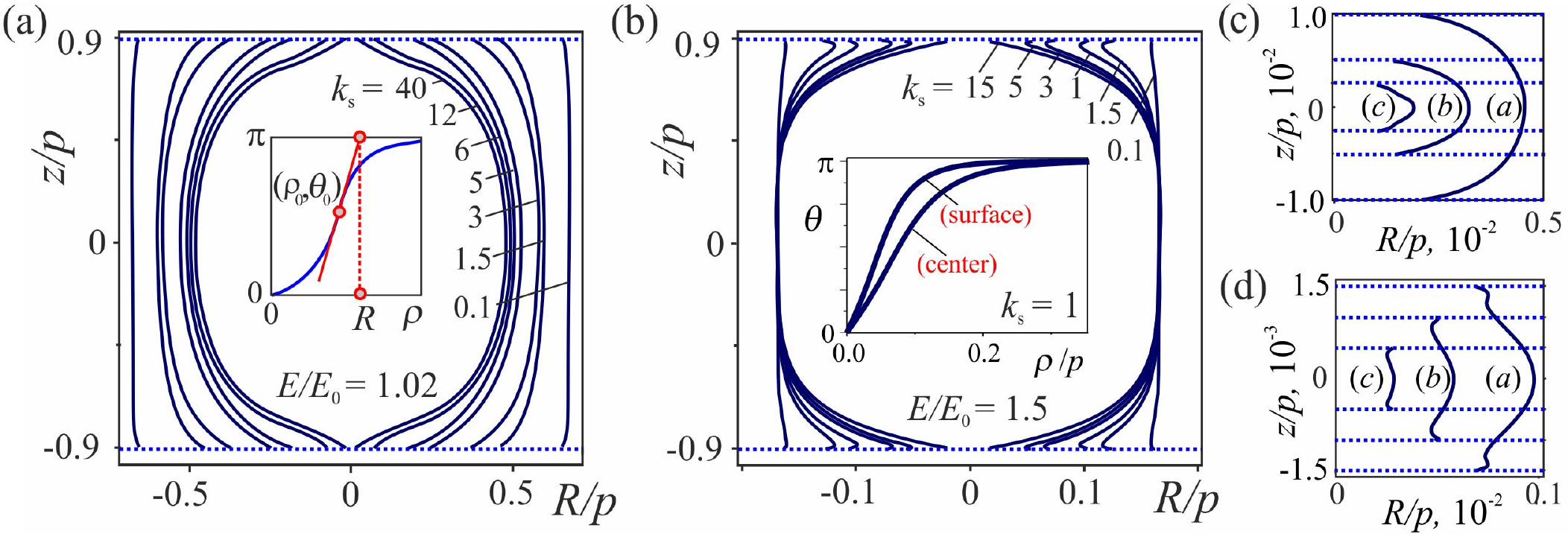}
\caption{
Equilibrium shapes $R (z)$ of spherulites in a layer with thickness $L = 1.8 p$
and for different values of homeotropic anchoring:
 (a) $E = 1.02 E_0$. Near the unwinding transition
spherulites have extended sizes; (b) $E = 1.5 E_0$. For
higher fields  their cores  become strongly localized.
Inset in fig. (a) introduces the effective  sizes of profiles $\theta (\rho)$
(Eq. \ref{width}); inset in fig. (b) shows spherulites shapes in layers
with thickness $L/p = 0.01 (\mathit{a}), 0.005 (\mathit{b}), 0.0025  (\mathit{c})$ 
( $E = 1.02 E_0$, $K_s = 7.0 K_{s0}$).
\label{sizes}
}
\end{figure*}
The boundary value problem (\ref{EulerSp1}), (\ref{EulerSp2}) 
has been solved by a standard finite-difference method with discretization 
on rectangular grids with adjustable grid spacings.
As initial guess for the iterative procedure by a Seidel method 
with Chebishev acceleration \cite{Press07} we used the known solutions 
of Eq. (\ref{EulerSp1}) for bulk chiral systems \cite{JMMM1994}, 
as starting profiles.
Solutions of Eqs. (\ref{EulerSp1}), (\ref{EulerSp2})   $\theta(\rho, z)$ 
can be presented as a stack of profiles $\theta (\rho)$ parametrized  
by $z$  ($ -L/2  <z < L/2$) (Fig. \ref{profiles}).
Under the influence of the homeotropic anchoring  $\theta (\rho)$  
curves vary along the layer thickness  from wild bell-shape lines 
in the center of the layer ($z = 0$) to narrow profiles  at the layer 
surface
($ z = \pm L/2$).
The equilibrium solutions for isolated axisymmetric skyrmions 
$\theta (\rho, z)$ strongly depend
on the control parameters: $\nu = L/p$, $ E/E_0$, and $K_s/K_{s0}$ 
(Figs. \ref{profiles}, \ref{sizes}).
They exist only for applied fields higher than the unwinding field ($ E > E_0$). 
Below this field axisymmetric skyrmions are unstable with respect to elliptic 
distortions (similar instabilities arise
in bulk chiral skyrmions \cite{pss1994}).
Formally the solutions for axisymmetric skyrmions with well-defined sizes exist at arbitrary high
fields $E > E_0$. However, the solutions with extended cores arise only for applied fields 
in the close proximity of the transition field $E \geq E_0$ (Figs. \ref{profiles}, \ref{sizes}).
With increasing field the core size rapidly decreases to the values comparable with the molecular length
manifesting a breakdown of the continuum theory.

For a $\theta (\rho)$ line the point  where the tangent at the inflection point ($\rho_0$, $\theta_0$)
intersects the $\rho$-axis (Fig. \ref{sizes}) introduces the radius $R (z)$ which characterizes the profiles width \cite{JMMM1994,Hubert1998}
 \begin{equation}
R = \rho_0 - \theta_0 \left(d \theta/d \rho \right)^{-1}_{\rho =\rho_0}
\label{width}
\end{equation}
Note that  a similar procedure is applied to introduce characteristic sizes 
for isolated domain walls and other solitonic states \cite{Hubert1998,pss1994}.
The calculated lines $R(z)$ ($-L/2 < z < L/2$) for different values of the control
parameters are plotted in Fig. \ref{sizes}.
They have a characteristic  convex shape.
The homeotropic surface anchoring compresses  ideally cylindrical axisymmetric solitons
into a convex-shaped spherulites (Figs. \ref{sizes}). However, a complex interplay between
surface volume interactions leads to other specific effects.
It was found that in an extend range of the control parameters the functions $R(z)$ reach
the mininum at a certain distance from the surface creating specific ``necks'' (Fig. \ref{sizes} (b)). 

%
\section{ Linear ansatz for isolated spherulites}
%
%
In a wide range of the control parameters solutions of Eqs. (\ref{EulerSp1}), (\ref{EulerSp2}) \ $\theta (\rho, z)$ are composed of arrow-like profiles $\theta (\rho)$ and can be satisfactorily approached by
a linear ansatz
\begin{equation}
\theta (\rho, z) = \pi \rho/\xi (z) 
\label{ansatz1}
\end{equation}
for $\rho \leq  \xi (z))$, and  $\theta (\rho, z) =  0$ for $\rho >  \xi (z)$.
A trial function (\ref{ansatz1}) is a specific case of a more general ``scaling ansatz'' 
\begin{equation}
\theta (\rho, z) = \theta (\rho/\xi(z)) 
\label{ansatz2}
\end{equation}
investigated for functional (\ref{energy1}) in \cite{JETPL2000}.

With ansatz (\ref{ansatz1})  energy $F$ (\ref{energy1}) can be reduced to
the following functional 
$\widetilde{F} (z) = (\pi K/4) \int_{-L/2}^{L/2} [\tilde{w} + \tilde{w}_s] dz$
with
\begin{align}
\tilde{w} (z) = \left(\frac{d \xi}{d z} \right)^2
+ 4\pi q_0 \xi + \frac{2 \varepsilon_a E^2}{K} \xi^2,
\label{functional3}
\end{align}
and $\tilde{w}_s = (K_s/K)^2 \xi^2  \delta (z \pm L/2)$.
Minimization of funtional $\widetilde{F} (z)$ yields the equation
\begin{align}
\xi (z)  = \frac{4 p}{\pi^2} \left(\frac{E}{E_0} \right)^{-2} 
\left[1 - \frac{1}{\Omega (\nu, k_s) }\cosh \left(\frac{\alpha z}{p} \right) \right],
\label{catenary}
\end{align}
describing a \textit{catenary} curve. 
Here $ \alpha = \pi (E/E_0) $  and
\begin{align}
\Omega(\nu, k_s) =  \cosh \left(\frac{\alpha \nu}{2} \right) 
+ \frac{\pi^2 k_s}{2}  \sinh \left(\frac{\alpha \nu}{2} \right) .
\label{omega}
\end{align}
The control parameters ($\nu$, $k_s$, $E/E_0$) (\ref{parameters}) determine
curvature and other parameters of line (\ref{catenary}) 
(for details see \cite{JETPL2000}).
Comparsion with solutions of Eqs. (\ref{EulerSp1}),
 (\ref{EulerSp2}) shows  that calculations with linear ansatz  (\ref{ansatz1}) 
reach a satisfactory quantitative accuracy only for weak anchoring ($k_s \ll 1$). 
Nevertheless analytical results for model (\ref{functional3})
offer an important insight into physics of confined chiral skyrmions.

The equilibrium sizes of  spherulites are formed as a result of a competition 
between the terms linear and quadratic with respect to $\xi$ in functional 
(\ref{functional3}). 
The former ($ \propto q_0$) is stemmed from the \textit{chiral} interactions 
imposed by the handedness of the system and which are represented in Frank
functional with energy contributions linear with respect to the first
spatial derivatives:
\begin{align}
w_q (\mathbf{n}) = 2K_2 q_0 \mathbf{n} \cdot \mathrm{rot} \mathbf{n}.
\label{bak}
\end{align}
The latter includes internal interactions independent 
on spatial derivatives of $\textbf{n}$ 
(as the dielectric anisotropy energy $\propto E^2$ in functional (\ref{density1}).
Finally, the first term in (\ref{functional3}) determines a variation of the equilibrium
core size $\xi (z)$ along the layer thickness imposed by the surface anchoring.
Note, functional (\ref{functional3}) does not include the energy contributions 
quadratic in the spatial derivatives (\textit{spray-twist-band} elastic energy 
contributions in (\ref{Frank})). 
Solutions (\ref{catenary}) for functional  (\ref{functional3}) display in a simple
form the fundamental features of the chiral skyrmion energetics and elucidate a crucial
role of the chiral energy $w_q (\mathbf{n})$ in their formation.
Energy contribution $w_q (\mathbf{n})$ arises in chiral liquid crystals 
\cite{Kleman2003} and cubic noncentrosymmetric magnets \cite{JETPL1995,Bak1980}
and are composed of antisymmetric differential forms linear with respect to spatial
derivatives of the order parameter $\mathbf{l}$ (so callled Lifshitz invariants) 
\cite{Bak1980}
\begin{align}
\Lambda_{ij}^{(k)} = l_i \frac{\partial l_j}{\partial x_k} 
- l_j \frac{\partial l_i}{\partial x_k}
\label{lifshitz}
\end{align}
Energy functionals of noncentrosymmetric condensed-matter systems 
contain chiral energy contributions $w_{\mathrm{chiral}} (\mathbf{l})$
constructed of the combinations of differential forms (\ref{lifshitz}) 
complied with their symmetry  \cite{JETPL1995,Bak1980}.
Particularly for isotropic and cubic systems 
$w_{\mathrm{chiral}} (\mathbf{l})$
reduced to 
$w_{q} (\mathbf{l}) = \Lambda_{yx}^{(z)} 
+ \Lambda_{xz}^{(y)} + \Lambda_{zy}^{(z)}$ =
$\mathbf{l} \cdot \mathrm{rot} \mathbf{l}$ describing
chiral interactions in liquid crystals and 
noncentrosymmetric cubic magnets and ferroelectrics
(\ref{bak}) \cite{Nature2006,Kleman2003,Bak1980,Wright1989}.
Energy contributions composed of differential forms of type (\ref{lifshitz})
provide a specific stabilization mechanism for two- and three-dimensional
localized states \cite{JETPL1995}. 
\begin{figure}
\centering
\includegraphics[width=1.0\columnwidth]{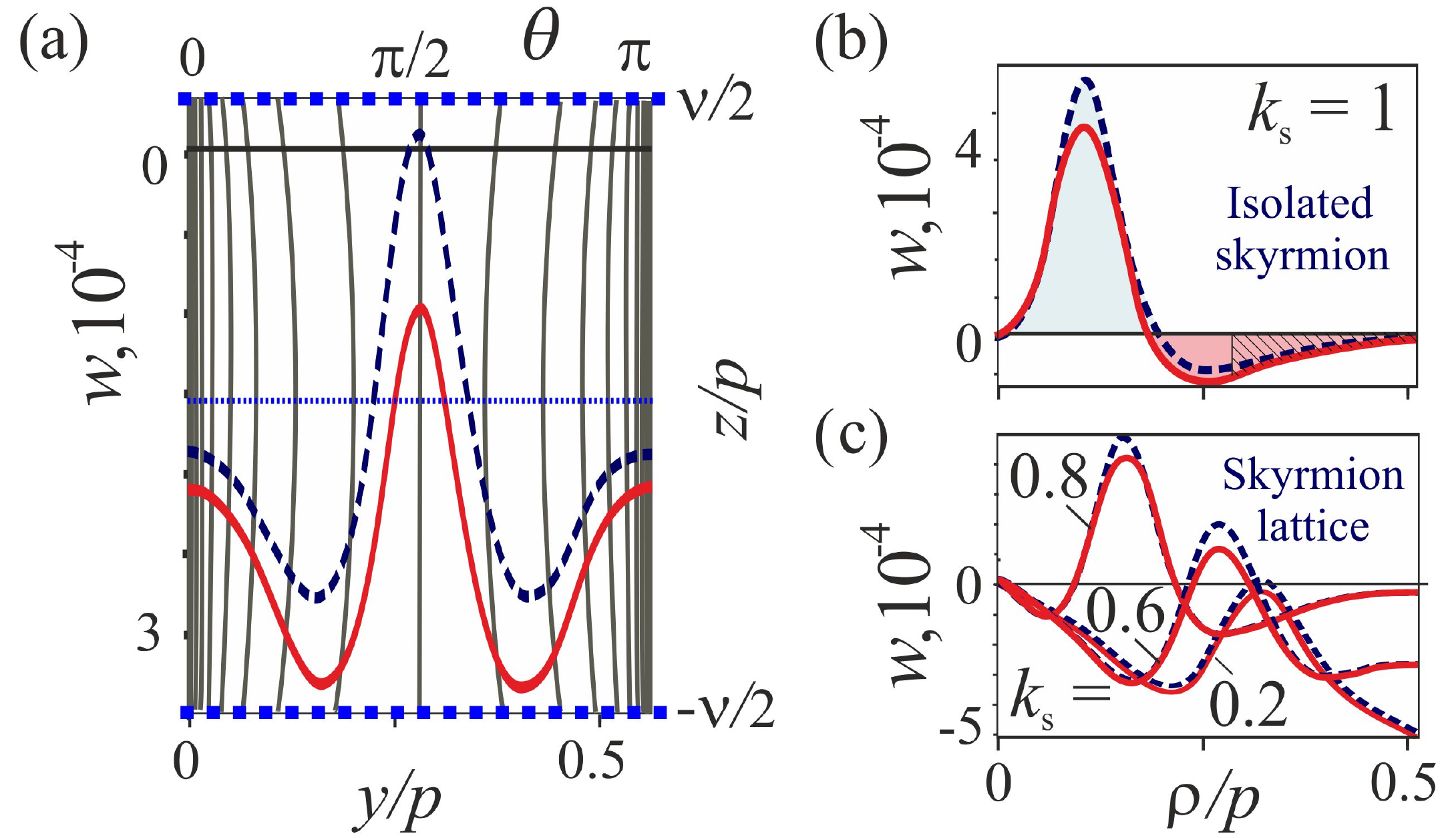}
\caption{(Color online) Typical solutions for
confined helicoids (a).
Top and right axes show profiles
$\theta (z)$ in equdistant planes $xz$.
Lower and left axes show energies densities:
dashed (blue) lines correspond to surface layers
($z = \pm L/2$) and the solid (red) line is for the
center of the layer ($z = 0$).
Corresponding energy densities for a isolated
sherulite and a skyrmion lattice are plotted
in figs. (b) and (c).
\label{helix2}
}
\end{figure}
\begin{figure}
\centering
\includegraphics[width=1.0\columnwidth]{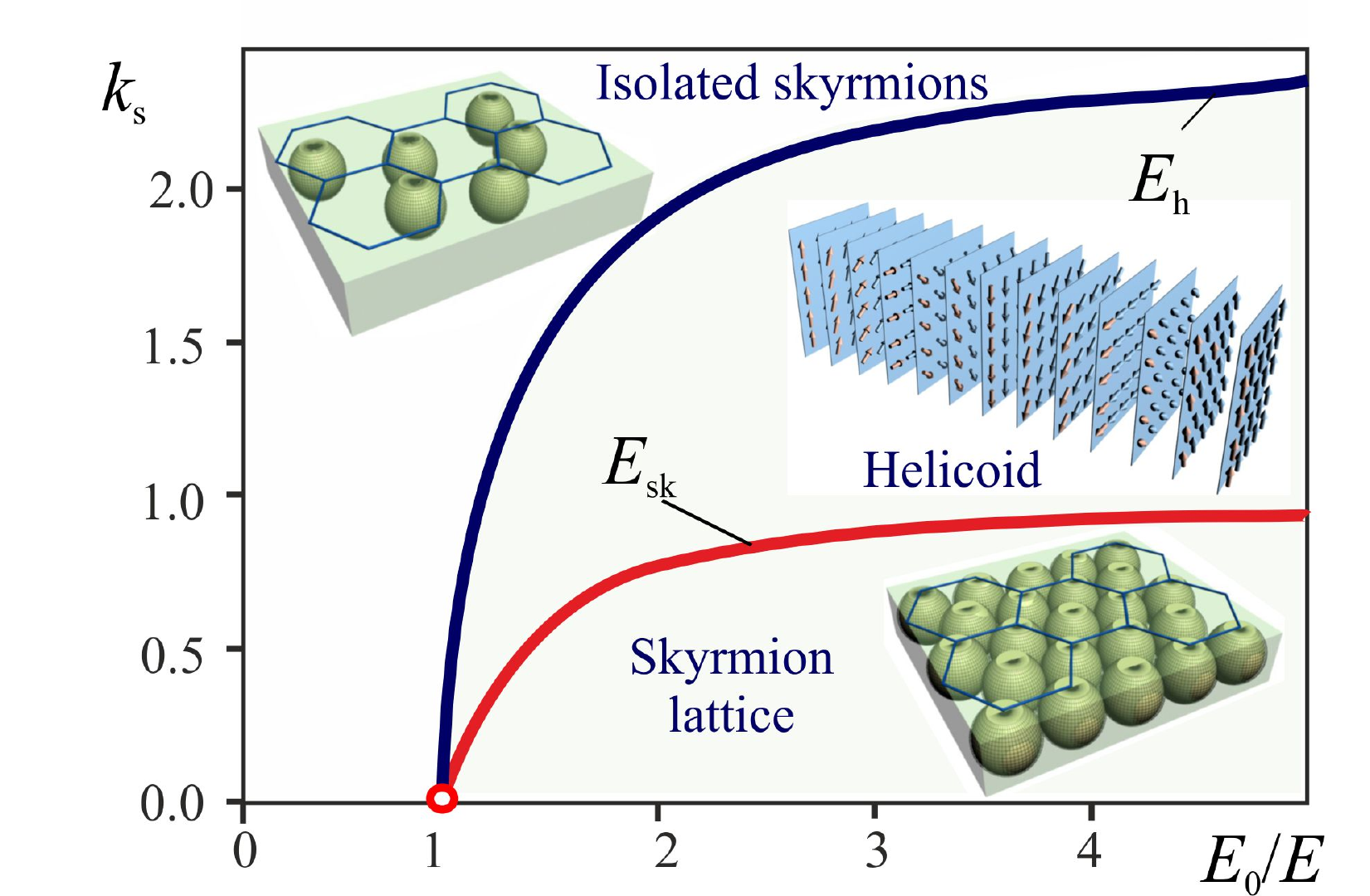}
\caption{(Color online)
The phase diagram of the equilibrium states
in reduced values of applied field ($E/E_0$)
and homeotropic anchoring ($k_s$) (\ref{parameters})
indicate the existence areas 
of the helicoid and the skyrmion lattice in a layer
with the confinement ratio $\nu = 1.8$.
\label{diagram}
}
\end{figure}
Importantly that classical skyrmions intensively investigated in
nonlinear physics are stabilized by higher order spatial derivatives 
of the order parameter (commonly referred to as \textit{Skyrme mechanism} 
\cite{skyrmion}).
Because in condensed-matter physics there are no physical interactions
described by higher order spatial derivatives (see e.g. Ref. 7 in
paper \cite{Butenko2010}), 
noncentrosymmetric condensed-matter systems (including chiral liquid crystals, 
multiferroics, magnetic systems with intrinsic and induced chirality) 
are of special importance as a particular class of materials where mesoscopic 
skyrmions can be created and manipulated.
This also attaches special importance to the solutions of 
Eqs. (\ref{EulerSp1}), (\ref{EulerSp2})  as basic elements 
providing the stability of axisymmetric skyrmions and other
solitonic states observed in chiral liquid crystals
layers \cite{Ackerman2014,Oswald2005}.

%
\section{ Confined  skyrmion lattices and helicoids}
%
%
In unconfined chiral liquid crystals solutions for axisymmetric skyrmions are homogeneous along 
their axes (solutions of type $\theta (\rho)$, $\psi = \pi/2 + \varphi$) \cite{JETP1998}. 
For $E < E_0$ (\ref{pitch}) the one-dimensional modulated states (\textit{helices})
correspond to the global miminum of the system \cite{Kleman2003}. Below critical field $E_0$ 
chiral skyrmions condense into \textit{metastable} lattices \cite{JETP1998}.
Near the critical field $E_0$ skyrmion lattices transform into honeycomb nets of thin 180 $^{\circ}$ walls. Helices in this region consist of broad stripes (with $\theta \approx0$) separated with thin
180 $^{\circ}$ walls. The equilibrium periods of the helices and skyrmion lattices tend to infinity as $E \rightarrow E_0$, and both modulated phases transform into the homogeneous state in the same critical point  (Fig. 10 in \cite{JETP1998}).
Contrary to bulk systems, in thin layers chiral modulations become inhomogeneous through the layer thickness. Here we consider main effects imposed by the homeotropic anchoring on helicoids and skyrmion lattices with propagation directions in the layer plane.
Within a circular-cell approximation (see e.g. \cite{JETP1998}) the equilibrium parameters of
a skyrmion lattice are derived by solving  Eq. (\ref{EulerSp1}) with the boundary conditions
$\theta(0,z)=0$, $\theta(a,z)=\pi$, (\ref{EulerSp2}) for different values of the cell size $a$
and minimization of the skyrmion lattice energy density with respect to $a$.
Similarly, the equilibrium states of a 2D helicoid propagating along the $y-$axis in the layer plane,
$\mathbf{n} = (\sin \theta (y, z),0, \cos \theta (y,z))$ are derived by solving equation
\begin{align}
\frac{\partial ^2 \theta}{\partial z^2} +\frac{\partial ^2 \theta}{\partial y^2} 
-\frac{\pi^2}{16} \left( \frac{E}{E_0} \right)^2 \sin\,\theta\,\cos\,\theta =0,
\label{spiral}
\end{align}
with the boundary conditions $\theta(0,z)=0$, $\theta(b,z)=2\pi$, (\ref{EulerSp2}) and optimization
of the helix energy density with respect to pitch $b$.
Typical solutions for Eq. (\ref{spiral}) are presented 
in Fig. \ref{helix2} (a) as $\theta(z)$  profiles in the equidistant $xz$ planes for a number of fixed values of $y$ (indicated with thin black lines ). 
%
Similarly to confined skyrmions  helix profiles $\theta (y, z)$ have a convex shape in the $xz$ planes.
%
%
The distribution of the energy density in the helicoid along the propagation direction $y$ is plotted in Fig. \ref{helix2} (a) for the surfaces $z =\pm L/2)$ (dashed blue line) and in the center of the layer $z = 0)$ (solid red line). The largest loss of the rotational energy occurs for the planes with $\theta=\pi/2$. 
Similar energy densities $w (\rho)$ for the isolated spherulite and the skyrmion lattice cell are plotted
in figs. (b) and (c).
%
%
The results of the calculations for confined helices and skyrmion lattices in a layer with
$\nu= 1.8 $ are presented in the phase diagram in reduced variables $E/E_0$ and $k_s$
 (Fig. \ref{diagram}). The confined helicoid is the global minimum of the system in the area
below critical line $E_h (k_s)$. A sufficiently strong homeotropic anchoring suppresses helical
modulations and above $E_h (k_s)$ line the homogeneous phase with $\theta =0$ has the lowest
energy. In this ``saturate'' phase isolated spherulite can exist as metastable states.
The skyrmion lattices can exist as metastable states below critical line $E_{sk} (k_s)$ and
trasform  into the homogeneous state at this critical line.
Below $E_h (k_s)$ line isolates spherulites are eliptically unstable (cf. \cite{pss1994}) and strip out into helicoids or into specific textures consisting of alongated 2D solitons 
(so called cholesteric fingers).
Experimental investigations of these modulated states are reviewed in \cite{Oswald2005}.

%
\section{ Comparison with magnetic chiral skyrmions}
%
%
Chiral magnetic skyrmions have been observed in nanolayers of cubic helimagnets 
\cite{Yu2010,Seki2012,Huang2012} and monolayers of
ferromagnetic metals with induced chiral interactions \cite{Romming2013}. 
It was established that surface/interface induced uniaxial anisotropy plays a
crucial role to stabilize skyrmions in these systems \cite{Huang2012,Romming2013,Wilson2014}.
Theoretically chiral modulations in nanolayers of chiral ferromagnets are described
by the energy density functional  \cite{Butenko2010,Huang2012,Wilson2014,Bak1980}
\begin{equation}
f_{\mathrm{m}}=A(\mathrm{grad}\,\mathbf{m})^2+D \mathbf{m}\cdot{\mathrm{rot}}\,\mathbf{m}
- \mathbf{H} \cdot \mathbf{M}- K_u (\mathbf{m} \cdot \mathbf{a})^2
\label{cubichelimagnet}
\end{equation}
which includes the exchange energy with coefficient $A$, the Dzyaloshinskii-Moriya coupling
($D$), the interaction with the applied magnetic field $\mathbf{M}$ (Zeeman energy), and
induced uniaxial anisotropy ($K_u$). $\mathbf{m} = \mathbf{M}/|\mathbf{M}$ is the unity vector along the magnetization  $\mathbf{M}$,  and $\mathbf{a}$ is the unity vector along the uniaxial anisotropy axis (along $z$ axis in this paper).
For chiral liquid crystals in applied electric fields  
the Frank free energy density functional in the one-constant approximation 
can be reduced to the following expression
\begin{equation}
f_{\mathrm{v}}=\frac{K}{2} (\mathrm{grad}\,\mathbf{n})^2
+Kq_0\mathbf{n}\cdot{\mathrm{rot}}\,\mathbf{n}
-\frac{\varepsilon_a}{2}(\mathbf{n}\cdot\mathbf{E})^2 .
\label{Frank2}
\end{equation}
We use here equation
$ (\mathrm{grad}\,\mathbf{n})^2 = (\rm{div}\,\mathbf{n})^2
+(\bf{n}\cdot\rm{rot}\,\mathbf{n})^2 +(\mathbf{n}\times\rm{rot}\,\mathbf{n})^2 $
 $+ <\mathrm{surface}$ $\mathrm{terms}>$ 
holding for any unity vector $\mathbf{n}$ (for details see e.g. \cite{Hubert1998}).
At zero field the energy density of a cubic helimagnet 
(\ref{cubichelimagnet})  has the same functional 
form with that of a chiral liquid crystal (\ref{Frank2}).
Thus, the solutions for skyrmions derived within model 
(\ref{cubichelimagnet}) \cite{JMMM1994,Butenko2010,Wilson2014}
at zero field describe  skyrmions in
bulk chiral liquid crystals at applied electric (magnetic) fields.
Also the energy functional for 
noncentrosymmetric antiferromagnets (Eq. (8) in \cite{PRB2002})
has a similar structure with functional  (\ref{Frank2}).
Physical relations between magnetic and liquid crystal skyrmionic states
have been discussed in \cite{JETP1998}.
It is known that surface/interface induced enhanced perpendicular uniaxial anisotropy arising in
nanolayers of magnetic metals  imposes a number of reorentation effects \cite{Fert1992}
which are similar to those induced by surface anchoring in liquid crystals \cite{Kleman2003}.
In existing epitaxial layers of cubic helimagnets  synthesized on Si(111) substrates
a uniaxial magnetic anisotropy is imposed by strains arising due to the lattice mismatch
between the magnetic layer and the substrate \cite{Karhu2010}.
In these nanolayers the induced anisotropy has a  volume-like character
(see Eq. (\ref{cubichelimagnet})). However, in nanolayers of noncentrosymmetric magnets
with induced anisotropy localized on their surfaces chiral skyrmions and helicoids are 
described by equations similar to those considered in this paper 
( Eqs. (\ref{EulerSp1}), (\ref{EulerSp2}), (\ref{spiral})).
Finally we mention that surface twisted modulations recently discovered 
in nanolayers of cubic helimagnets \cite{Meynell2014} are expected to occur
in confined chiral liquid crystals.
%
%
\section{ Conclusions.}
%
%
We present numerically exact solutions for isolated and 
embedded axisymmetric skyrmions in a thin layer of a chiral 
liquid crystal with homeotropic anchoring.
The interplay between the elastic stiffness, chiral twists, and perpendicular
surface leads to the formation of 2D skyrmions with a characteristic ``barrel''
(spherulite) shape (Fig. \ref{pics}).
The basic equations for confined axisymmetric skyrmions 
(Eqs. (\ref{EulerSp1}), (\ref{EulerSp2})) and other chiral
modulations depend on three
material parameters (\ref{parameters}).
Detailed analysis of these solutions in the full phase space
of the control parameters ($E/E_0, k_s, \nu$) will be done
elsewhere.
In this paper we restrict our analysis 
to a few representative samples 
(Figs. \ref{profiles}, \ref{sizes}, \ref{helix2})
illustrating the basic features of confined chiral modulations
 and demonstrating a fundamental role of axisymmetric
strings  (\ref{EulerSp1}), (\ref{EulerSp2})) in the formation
of two-dimensional solitonic states in thin layers of chiral liquid
crystals.
Multi-dimensional modulated textures observed in confined chiral
liquid crystals consist of 2D nonsingular axisymmetric strings 
(skyrmions proper) \cite{Ackerman2014} and  2D and 3D localized structures 
including different types of point defects and dislocations
\cite{Ackerman2014,Kawachi1974,Oswald2005,Smalyukh2005,Smalyukh2010,Chen2013,Pandey2014}.
The former are described by solutions of Eqs. 
(\ref{EulerSp1}), (\ref{EulerSp2}).
The latter includes nonsigular axisymmetric filaments as
basic stabilization elements.
Next efforts  on calculations of 2D and 3D chiral solitons  
including  point and linear singularities
should provide a theoretical basic for detailed analysis 
of  solitonic states arising in confined chiral liquid crystals.

\begin{acknowledgements}
The authors are thankful to J. Fukuda, T. Monchesky, M. Mostovoy, I. I. Smalyukh for useful discussions. A. O. L. acknowledges support from the Stichting voor Fundamenteel Onderzoek der Materie (FOM).
A.N.B. acknowledges financial support by DFG through Grant No. BO 4160/1-1.
\end{acknowledgements}


\begin{thebibliography} {99}


\bibitem{skyrmion} 
\textit{The Multifaceted Skyrmion}, edited by G. E. Brown and M. Rho
(World Scientific, Singapore,2010)

\bibitem{JETPL1995}
 A.  Bogdanov, Pis'ma Zh. Eksp. Teor. Fiz. \textbf{62}, 231 (1995) 
[JETP Lett. \textbf{86}, 247 (1995)];
A. N. Bogdanov, D. A. Yablonsky, Zh. Eksp. Teor. Fiz. \textbf{95}, 178 (1989) 
[JETP \textbf{68}, 101 (1989)]. 

\bibitem{JETP1998}
 A. N. Bogdanov,  A. A. Shestakov, 
Zh. Eksp. Teor. Fiz. \textbf{113}, 1675 (1998) 
[JETP \textbf{86}, 911 (1998)]; 
A. N. Bogdanov, U. K. R{\"o}{\ss}ler, A. A. Shestakov, 
Phys. Rev. E \textbf{67}, (2003).

\bibitem{JMMM1994} 
A. Bogdanov  and A. Hubert, 
J. Magn. Magn. Mater.\textbf{138}, 255 (1994); 
 \textbf{195}, 182 (1999).

\bibitem{Nature2006} 
U.K. R\"o\ss ler, A.N.\ Bogdanov, C. Pfleiderer,
Nature \textbf{442}, 797 (2006). 

\bibitem{Butenko2010}
A. B. Butenko et al., 
Phys. Rev. B {\textbf{82}}, 052403 (2010).

\bibitem{Borisov2010}
A. B. Borisov, F. N. Rybakov, 
Low Temp. Phys. \textbf{36}, 766 (2010).

\bibitem{JETPL2000}  
A. N. Bogdanov JETP Lett. \textbf{71}, 85 (2000).

\bibitem{Fukuda2011}
J. Fukuda, S. Zumer, Nature Comm. 
{\textbf{2}}, 246 (2011).

\bibitem{Muhlbauer2009}
S. Muhlbauer et al.,
Science, {\textbf{323}}, 915 (2009);
%
C. Pappas et al.,
Phys. Rev. Lett.  {\textbf{102}}, 197202 (2009).

\bibitem{Wilhelm2011}
H. Wilhelm et al.,
J. Phys.: Condens.Matter \textbf{24}, 294204 (2012).

\bibitem{Yu2010}
X. Z. Yu et al. 
Nature (London), {\textbf{465}}, 901 (2010).

\bibitem{Seki2012}
S. Seki et al.,
Science \textbf{336}, 198 (2012).
 
\bibitem{Huang2012}
S. X. Huang and  C. L. Chien, 
Phys. Rev. Lett. 
{\textbf{108}}, 267201 (2012);
%
M. N. Wilson, E. A. Karhu, A. S. Quigley et al., 
Phys. Rev. B 
{\textbf{86}}, 144420 (2012).

\bibitem{Romming2013}
S. Heinze et al., 
Nature Phys. {\textbf{7}}, 713 (2011);
N. Romming et al., 
Science {\textbf{341}}, 636 (2013).

\bibitem{JPCS2011} 
U.K. R\"o\ss ler, A. A. Leonov, A.N.\ Bogdanov, 
J.Phys. Conf. Ser. \textbf{303}, 012105 (2011). 

%
\bibitem{Wilson2014}
M. N. Wilson et al., 
Phys. Rev. B {\textbf{89}}, 094411 (2014).

\bibitem{Kiselev2011}
N. S. Kiselev et al. 
J. Phys. D \textbf{44}, 392001 (2011).

\bibitem{Fert2013}
A. Fert, V. Cros, and J. Sampaio, 
Nat. Nano \textbf{8}, 152 (2013).

\bibitem{Ackerman2014}
P. J. Ackerman et al., 
Phys. Rev. E  \textbf{90} , 012505 (2014).

\bibitem{Kawachi1974}
M. Kawachi, O. Kogure, Y. Kato,
Japan. J. Appl. Phys. \textbf{13}, 1457 (1974);
W. E. L. Haas, J. E. Adams,
Appl. Phys. Lett. \textbf{25}, 263 (1974).

\bibitem{Oswald2005}
P. Oswald, P. Pieranski, 
\textit{Nematic and Cholesteric Liquid Crystals},
(Taylor\&Francis Group, London, 2005).

\bibitem{Smalyukh2005}
I. I. Smalyukh et al., 
%
Phys. Rev. E, \textbf{72}, 061707 (2005).

\bibitem{Smalyukh2010}
I. I. Smalyukh et al., 
%
Nature Mater. \textbf{9} , 139 (2010).

\bibitem{Chen2013}
B. G. Chen et al., 
%
Phys. Rev. Lett., \textbf{110}, 237801 (2013).


\bibitem{Pandey2014}
M. B. Pandey et al., 
%
Phys. Rev. E,  \textbf{89}, 060502 (2014).


\bibitem{Derrick1964}
G. H. Derrick, J. Math. Phys. \textbf{5}, 1252 (1964).
%

\bibitem{Kleman2003}
P. G. De Gennes, J. Prost, 
\textit{The Physics of Liquid Crystals} 
(Oxford University Press, Oxford, 1993);
%
M. Kleman, O. D. Lavrentovich,
\textit{Soft matter physics: an introduction},
(Springer-Vertag, New York, 2003).



\bibitem{Akahane1976}
T. Akahane, T. Tako
Japan. J. Appl. Phys. \textbf{15}, 1559 (1976).

\bibitem{pss1994} 
A. Bogdanov  and A. Hubert, 
phys. stat. sol. (b) \textbf{186}, 527 (1994).


\bibitem{Press07}  W. H. Press, S. A. Teukolsky, 
W. T. Vetterling, B. P. Flannery, 
\textit{Numerical Recipes - 
The Art of Scientific Computing}
 (Cambridge University Press, Cambridge 2007). 

 \bibitem{Hubert1998}
A. Hubert, R. Sch\"{a}fer, \textit{Magnetic Domains} 
(Springer, Berlin, 1998).


\bibitem{Bak1980}  
I.\ E.\ Dzyaloshinskii, 
Sov.\ Phys.\ JETP {\textbf{20}}, 665 (1964);
P.\ Bak and M.\ H.\ Jensen, 
J.\ Phys. C: Solid State Phys.
\ {\textbf 13}, L881 (1980).

\bibitem{PRB2002}
A. N. Bogdanov, U. K. R\"o\ss ler, M. Wolf, and  K. H. M\"uller,
Phys. Rev. B {\textbf{66}}, 214410 (2002).

\bibitem{Fert1992}
A. Thiaville and A. Fert
J. Magn. Magn. Mater. {\textbf{113}}, 161 (1992);
A. N. Bogdanov, U. K. R\"o\ss ler,  K.H. M\"uller,
J. Magn. Magn. Mater. {\textbf{238}}, 155 (2002).


\bibitem{Karhu2010}
E. A. Karhu et al.,
Phys. Rev. B {\textbf{82}}, 184417 (2010);
E. A. Karhu et al.,
Phys. Rev. B {\textbf{85}}, 094429 (2012).

\bibitem{Meynell2014} 
S. A. Meynell et al. 
Phys. Rev. B {\textbf{90}}, 014406 (2014).


\bibitem{Wright1989}
D. C Wright, N. D. Mermin, 
Rev. Mod. Phys. \textbf{61}, 385 (1989).





\end{thebibliography}
\end{document}